\documentclass[aps,prb,twocolumn,showpacs,floatfix,amsmath]{revtex4}

\usepackage{graphicx}
\begin{document}


\title{Tunneling spectra of submicron
Bi$_2$Sr$_2$CaCu$_2$O$_{8+\delta}$ intrinsic Josephson junctions:
Evolution from superconducting gap to pseudogap}

\author{S. P. Zhao, X. B. Zhu, and H. Tang}

\affiliation{Beijing National Laboratory for Condensed Matter
Physics, Institute of Physics, Chinese Academy of Sciences, Beijing
100190, China}


\begin{abstract}

Tunneling spectra of near optimally doped, submicron
Bi$_2$Sr$_2$CaCu$_2$O$_{8+\delta}$ intrinsic Josephson junctions are
presented, and examined in the region where the superconducting gap
evolves into pseudogap. The spectra are analyzed using a self-energy
model, proposed by Norman {\it et al.}, in which both quasiparticle
scattering rate $\Gamma$ and pair decay rate $\Gamma_{\Delta}$ are
considered. The density of states derived from the model has the
familiar Dynes' form with a simple replacement of $\Gamma$ by
$\gamma_+$ = ($\Gamma$+$\Gamma_{\Delta}$)/2. The $\gamma_+$
parameter obtained from fitting the experimental spectra shows a
roughly linear temperature dependence, which puts a strong
constraint on the relation between $\Gamma$ and $\Gamma_{\Delta}$.
We discuss and compare the Fermi arc behavior in the pseudogap phase
from the tunneling and angle-resolved photoemission spectroscopy
experiments. Our results indicate an excellent agreement between the
two experiments, which is in favor of the precursor pairing view of
the pseudogap.

\end{abstract}


\maketitle

\section{Introduction}

Pseudogap in cuprate superconductors is now a well established
phenomenon, \cite{timu99} but its nature remains a mystery. There
are theories that provide viable candidates for the origin of the
pseudogap. Some of them involve precursor superconductivity,
\cite{emer95,norm98a} such as those based on boson-mediated pairing
\cite{norm06} or bipolaron \cite{alex00,msb07} scenario, or on the
resonating valence bond state. \cite{lee06} Other theories like the
nearly antiferromagnetic Fermi liquid model \cite{pines97} have also
been proposed. Moreover, pseudogap can arise from a different origin
such as charge density wave that competes with the superconducting
state. \cite{li06} Apparently, understanding pseudogap is crucial
for finding the underlying pairing mechanism of the cuprate
superconductors.

One important observation in the pseudogap phase from angle-resolved
photoemission spectroscopy (ARPES) experiments is the existence of
an ungapped portion on the Fermi surface, described through the
spectral function $A({\mathbf k},\omega)$, known as the Fermi arc.
\cite{norm98b} Recently, there has been considerable progress
towards its understanding. ARPES measurements on
Bi$_2$Sr$_2$CaCu$_2$O$_{8+\delta}$ (Bi-2212) materials demonstrated
a linear temperature dependence of arc length above $T_c$
\cite{kani06} and the opening of a gap on the arc below $T_c$.
\cite{shen07,kani07} Further analysis of the ARPES,
\cite{norm07,chub07} Raman, and specific heat data \cite{stor07}
showed that the arc is interpretable from a linear temperature
dependence of the quasiparticle scattering rate $\Gamma$.

Tunneling has traditionally been a useful tool in revealing the
material's superconducting properties. \cite{schr64,wolf85} For the
Bi-2212 materials, experimental techniques including scanning
tunneling microscope (STM), \cite{renn98} break junctions (BJs),
\cite{zasa} and intrinsic Josephson junctions (IJJs)
\cite{kras00,suzu99,bae08,zhu06,zhu07,zhao07} were used, yielding a
variety of valuable results. For instance, Zasadzinski {\it et al.}
analyzed the BJ and STM data using the $d$-wave Eliashberg theory
and correlated the tunneling dip feature with the magnetic resonance
mode. \cite{zasa} These studies point to the electron coupling to a
narrow boson spectrum as possible pairing mechanism. From the STM
results, \cite{renn98} the pseudogap opening temperature
$T^{\star}$, rather than $T_c$, was suggested to be the mean-field
critical temperature, thus supporting precursor pairing as the
origin of the pseudogap phase. Meanwhile, in the IJJ studies, a
different observation, namely the coexistence of the superconducting
gap and pseudogap, which leads to the conclusion that their origins
are different, was reported. \cite{kras00,suzu99,bae08} Viewing from
these results, it is clear that the answer to the pseudogap nature
is still ambiguous from the tunneling experiments using different
techniques. Further understanding is therefore needed, in particular
on the above-mentioned Fermi arc issue that has not been addressed
so far in the reported tunneling experiments.

In this paper, we present and examine the tunneling spectra of near
optimally doped, submicron Bi-2212 IJJs near and above $T_c$ where
the superconducting gap evolves into pseudogap. \cite{rem} We show
that the spectra can be well fitted using the density of states
(DOS) in a form often used in the tunneling experiment:
\cite{dynes78}
\begin{equation}
\label{dynes} N(\theta,\omega) = \mbox{Re}\left[\frac{\omega
+i\gamma_+} {\sqrt{(\omega
+i\gamma_+)^2-\Delta^2\cos^2(2\theta)}}\right]~,
\end{equation}
\noindent where a $d$-wave gap is considered and $\theta$ is the
angle of in-plane momentum measured from ($\pi$,0). Such fit leads
to a slowly decreasing $\Delta$ and a roughly linear behavior of
$\gamma_+$ with increasing temperature. It can be easily shown that
Eq.~(1) is closely related to the phenomenological self-energy:
\cite{norm98a}
\begin{equation}
\label{norman} \Sigma({\mathbf
k},\omega)=-i\Gamma+\frac{\Delta_{\mathbf
k}^2}{\omega+\epsilon_{\mathbf k}+i\Gamma_{\Delta}}~,
\end{equation}
\noindent where $\epsilon_{\mathbf k}$ is the energy of bare
electrons relative to the value at the Fermi surface and
$\Gamma_{\Delta}$ comes from the pair fluctuations. The self-energy
in Eq.~(2), applicable for the description of precursor
superconductivity, was used recently to discuss the Fermi arcs from
ARPES experiment assuming $\Gamma$ = $\Gamma_{\Delta}$.
\cite{norm07,chub07} In the present work, we will consider the more
general case where $\Gamma$ $\not=$ $\Gamma_{\Delta}$ based on the
experimentally obtained tunneling parameters, and compare the data
with the ARPES experiment. As will be seen below, our results
demonstrate a remarkable consistency between the two experiments,
which is in favor of the precursor pairing view of the pseudogap.

\section{Tunneling spectra of submicron IJJs near optimal doping}

\begin{figure}[t]
\includegraphics[width=0.42\textwidth]{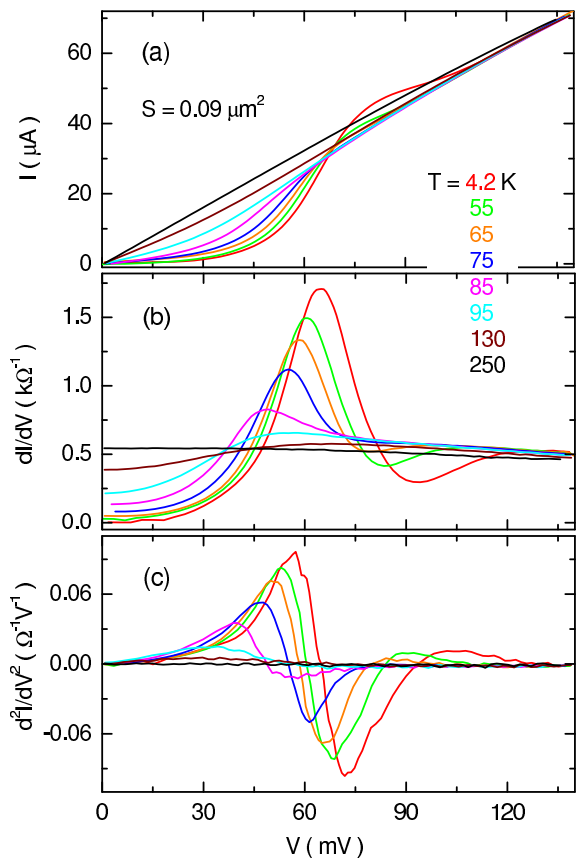}
\caption{(Color online) (a) $I$-$V$, (b) $dI/dV$, and (c)
$d^2I/dV^2$ characteristics of a near optimally doped, submicron
Bi-2212 mesa ($T_c$ = 89 K) containing 10 IJJs. $V$ corresponds to
the voltage per IJJ.} \label{fig:expiv}
\end{figure}

\begin{figure}[t]
\includegraphics[width=0.42\textwidth]{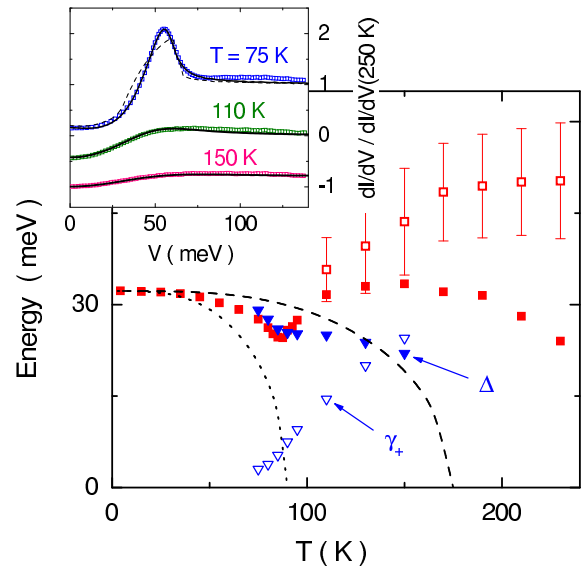}
\caption{(Color online) Half the $dI/dV$ peak energy $\Delta_p$
(solid squares). Open squares are those from the spectra normalized
to the one at 250 K. Solid and open triangles are $\Delta$ and
$\gamma_+$ in Eq.~(1) from fitting the spectra using
$N_{eff}(\omega)$. The dashed and dotted lines are the BCS $d$-wave
$\Delta_p(T)$ with $2\Delta_p(0)/kT_c$ = 4.28 and that with
temperature normalized to $T_c$, respectively. The inset shows the
normalized spectra at three temperatures (symbols), and the
calculated results using $N_{eff}$ (solid lines) and $N_d(\omega)$
(dashed line at 75 K). The 110 and 150 K curves are shifted
downwards for clarity.} \label{fig:didvtc}
\end{figure}

IJJs, being intrinsic, have the advantage of avoiding the problems
like the surface deterioration and unstable junction structure, and
can offer a convenient temperature or magnetic-field dependent
measurement. When the IJJ size decreases, the sample's self-heating,
a main obstacle in the IJJ studies, will also decrease.
\cite{kras01} Recently, we showed that self-heating can be reduced
significantly when junction sizes decrease down to the submicron
level. \cite{zhu06,zhu07} In Fig.~\ref{fig:expiv}, we plot the
tunneling spectra of an IJJ mesa, 0.3 $\mu$m in size, fabricated on
a Bi-2212 crystal with $T_c$ = 89 K. The spectra show a well-defined
normal-state resistance $R_N$ $\sim$ 1.92 k$\Omega$, and clear
low-$T$ dip feature \cite{rem1} as observed in the BJ experiment.
\cite{zasa} It is seen that as temperature increases, the
superconducting peak shifts to lower voltages and its strength
weakens. Near $T_c$, the superconducting gap smoothly evolves into
pseudogap that persists up to temperatures as high as 230 K, similar
to the STM results. \cite{renn98} In Fig.~\ref{fig:didvtc}, we plot
half the $dI/dV$ peak energy $\Delta_p$ as solid squares, while the
open squares represent the corresponding data for the spectra
normalized to the one at 250 K. Both data show a slope change near
150 K, which is reminiscent of the experiments suggesting two
pseudogap temperatures $T_1^{\star}$ and $T_2^{\star}$ (230 and 150
K, respectively, in the present case), such as Nernst experiment.
\cite{lee06,yayu05}

In Fig.~\ref{fig:expiv}, the $dI/dV$ dip disappears at $\sim$ 70 K,
which allows simple $I$-$V$ curve calculations \cite{wolf85} to fit
the data near and above $T_c$ using a $\theta$-averaged DOS:
\cite{zasa}
\begin{eqnarray}
N_d(\omega) = (1/2\pi)
\int\limits_{0}^{2\pi}N(\theta,\omega)d\theta~,
\end{eqnarray}
\noindent or an effective DOS:
\begin{eqnarray}
N_{eff}(\omega) = (1/\pi) \int\limits_{0}^{2\pi}N(\theta,\omega)
\cos^2(2\theta)d\theta~,
\end{eqnarray}
\noindent in which tunneling with a directional preference is
considered. \cite{zhao07,ozyu00} In Eqs.~(3) and (4),
$N(\theta,\omega)$ is given by Eq.~(1), which has the Dynes' form
\cite{dynes78} and is used frequently in fitting the experimental
tunneling spectra for both BCS \cite{wolf85} and cuprate
\cite{zasa,ozyu00} superconductors. In the inset of
Fig.~\ref{fig:didvtc}, we show the normalized experimental spectra
at three temperatures (symbols) together with the calculated results
(lines). From the data at 75 K, it can be seen that the result using
$N_{eff}(\omega)$ (solid line) fits the experimental spectrum much
better than that using $N_d(\omega)$ (dashed line). The fitting
parameters $\Delta$ and $\gamma_+$ using $N_{eff}(\omega)$ are
plotted in Fig.~\ref{fig:didvtc} as solid and open triangles,
respectively. The results show that $\Delta$ has a weak temperature
dependence while $\gamma_+$ increases roughly linearly with
temperature. At about $T_2^{\star}$ = 150 K, $\gamma_+$ becomes
larger than $\Delta$.

\section{Basic formulas}

Eq.~(1) can be derived from the self-energy in Eq.~(2) under certain
appropriate approximations, which can therefore be used as a
starting point to compare the tunneling and ARPES experiments. First
of all, we note that Fermi arcs in the ARPES experiment are usually
discussed through the spectral function $A({\mathbf k},\omega)$ on
the Fermi surface. It is considered to be gapped if $A(k_F,\omega)$
has maxima at $\omega$ = $\pm\omega_p$ $\not=$ 0 (we call it ARPES
gap below for convenience), while Fermi arc appears at places where
$A(k_F,\omega)$ has maximum only at $\omega$ = 0. From the
self-energy in Eq.~(2), it can be easily shown that the Green's
function $G({\mathbf k},\omega)$ = 1/$[\omega-\epsilon_{\mathbf
k}-\Sigma({\mathbf k},\omega)]$ has the form

\begin{equation}
G({\mathbf k},\omega)=
\frac{\omega+i\Gamma_{\Delta}+\epsilon_{\mathbf k}}
{(\omega+i\Gamma-\epsilon_{\mathbf k})
(\omega+i\Gamma_{\Delta}+\epsilon_{\mathbf k})-\Delta_{\mathbf
k}^2}~,
\end{equation}

\noindent therefore $A(k_F,\omega)$ = -(1/$\pi$)Im$G(k_F,\omega)$,
assuming $\Delta_{\mathbf k}$ = $\Delta\cos(2\theta)$, is given by

\begin{equation}
A(\theta,\omega)=\frac{1}{\pi}\frac{\Gamma_{\Delta}[\Delta^2\cos^2(2\theta)+\Gamma
\Gamma_{\Delta}] +\Gamma \omega^2}
{[\omega^2-\Delta^2\cos^2(2\theta)-\Gamma
\Gamma_{\Delta}]^2+\omega^2(\Gamma+\Gamma_{\Delta})^2}~.
\end{equation}

In a special case of $\Gamma$ = $\Gamma_{\Delta}$, the above
equation becomes
\begin{equation}
A(\theta,\omega)=\frac{\Gamma}{\pi}\frac{\omega^2+\Delta^2\cos^2(2\theta)+\Gamma^2}
{[\omega^2-\Delta^2\cos^2(2\theta)-\Gamma^2]^2+4\omega^2\Gamma^2}~,
\end{equation}
which was used by Norman {\it et al.} \cite{norm07} and Chubukov
{\it et al.} \cite{chub07} to interpret Fermi arcs in terms of the
$\Gamma$ parameter. Tunneling experiments, on the other hand, are
discussed through DOS, which is defined by -(1/$\pi$)
$\sum\limits_{\mathbf k}\mbox{Im}G({\mathbf k},\omega)$. In
Eliashberg's strong-coupling theory of BCS superconductors, the
summation over ${\mathbf k}$ is safely replaced by an integration
$\int\limits_{-\infty}^{\infty}d\epsilon_{\mathbf k}$ and the
strong-coupling DOS with a complex gap function $\Delta(\omega)$ can
be obtained. \cite{schr64} In the present case for the Bi-2212
materials, we can follow the same procedure if we assume a circular
Fermi surface \cite{chub07} and $\Delta_{\mathbf k}$ =
$\Delta\cos(2\theta)$. This is possible since, considering a
tight-binding dispersion like $\alpha$+2$\beta$(cos$k_x$+cos$k_y$)
where $\alpha$ and $\beta$ are constants, the DOS in the {\it
normal} state has a slow variation with energy for all $\theta$. A
simple derivation, with the introduction of
\begin{equation}
\gamma_{\pm}=\frac{\Gamma \pm \Gamma_{\Delta}}{2}~,
\end{equation}
shows that Eq.~(2) directly leads to Eq.~(1). In the case of
$\Gamma$ = $\Gamma_{\Delta}$, it becomes
\begin{equation}
\label{dynes} N(\theta,\omega) = \mbox{Re}\left[\frac{\omega
+i\Gamma} {\sqrt{(\omega
+i\Gamma)^2-\Delta^2\cos^2(2\theta)}}\right]~,
\end{equation}
which is the familiar Dynes' DOS phenomenologically proposed over
three decades ago. \cite{dynes78}

\section{Visualizing Fermi arcs using the tunneling parameters}

The $\gamma_+$ parameter in Fig.~\ref{fig:didvtc}, obtained from
fitting the experimental spectra and defined by Eq.~(8), provides a
strong constraint on the relation between $\Gamma$ and
$\Gamma_{\Delta}$. The $\Gamma_{\Delta}$ parameter, representing the
Cooper pair decay rate, has been discussed previously for the BCS
\cite{cohen69,tink96} and cuprate \cite{norm98a} superconductors. In
the case of BCS superconductors, it was taken into account to
explain the fluctuation effect above $T_c$ in materials such as
aluminum, where a DOS proposed by de Gennes for gapless
superconductivity has been used. \cite{cohen69} From a general
consideration, $\Gamma_{\Delta}$ should be zero below $T_c$ and is
known to have a linear temperature dependence above it:
\cite{cohen69,tink96}

\begin{equation}
\label{GDT} \Gamma_{\Delta}=\left\{\begin{array}{cc}
~0~,~~~~~~~~~~~~~~~~~ for~~~T < T_c~, \\
\\ \frac{8}{\pi}k_B(T-T_c)~,~~~ for~~~T \geq T_c~.
\end{array}
\right.
\end{equation}

\noindent The $\Gamma_{\Delta}$ behavior was also discussed in the
case of cuprate superconductors. In fitting the ARPES data of
underdoped Bi-2212 material with $T_c$ $\sim$ 83 K, the linear
temperature dependence of $\Gamma_{\Delta}$ was confirmed, but the
values were twice as large as the BCS weak-coupling case predicted
in Eq.~(\ref{GDT}).

\begin{figure}[t]
\includegraphics[width=0.42\textwidth]{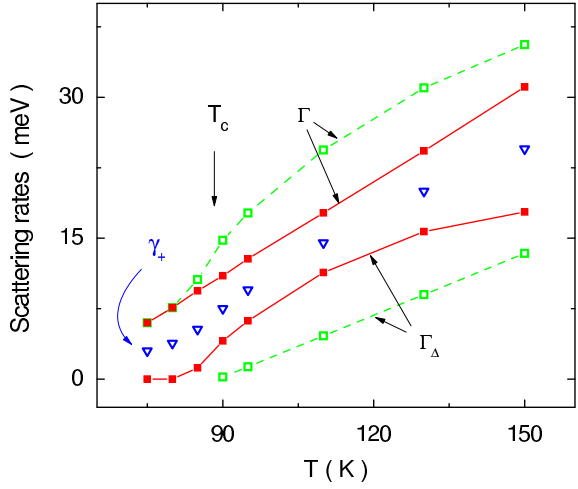}
\caption{(Color online) Two sets of the scattering rates $\Gamma$
and $\Gamma_{\Delta}$ obtained from experimentally determined
$\gamma_+$ parameter using Eq.~(8), assuming $\Gamma_{\Delta}$ in
Eq.~(10) (dashed lines with open symbols), and a $\Gamma$ linearly
increasing with temperature starting from below $T_c$ (solid lines
with full symbols).} \label{fig:rates}
\end{figure}

\begin{figure}[t]
\includegraphics[width=0.45\textwidth]{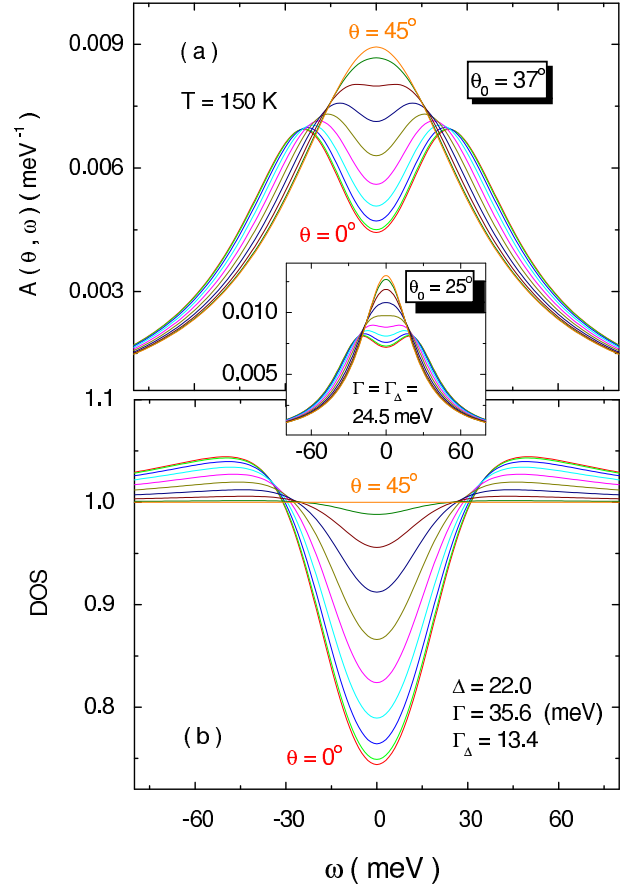}
\caption{(Color online) Spectral function $A(\theta,\omega)$ on the
Fermi surface at 150 K calculated from Eq.~(6) using $\Delta$,
$\Gamma$, and $\Gamma_{\Delta}$ parameters indicated. Curves are
plotted with $\theta$ step of 5$^{\circ}$. (b) Corresponding results
of DOS from Eq.~(1). The $\Gamma$ and $\Gamma_{\Delta}$ parameters
are taken from the dashed lines with open symbols at 150 K in
Fig.~\ref{fig:rates}. The inset shows $A(\theta,\omega)$ with equal
$\Gamma$ and $\Gamma_{\Delta}$ at the same temperature.}
\label{fig:spectral}
\end{figure}

For known $\Gamma_{\Delta}$, the $\Gamma$ parameter can be found
from Eq.~(8) with experimentally determined $\gamma_+$ values in
Fig.~\ref{fig:didvtc}, and the spectral function $A(\theta,\omega)$
can be calculated from Eq.~(6). The dashed lines with open symbols
in Fig.~\ref{fig:rates} are such evaluated $\Gamma_{\Delta}$ and
$\Gamma$ parameters using Eq.~(10). The $\Gamma$ parameter thus
obtained shows an enhanced and clear nonlinearity in its temperature
dependence, while $\gamma_+$ exhibits only a small nonlinearity. In
Fig.~\ref{fig:spectral}(a), we show the calculated results of
$A(\theta,\omega)$ using the $\Delta$, $\Gamma$, and
$\Gamma_{\Delta}$ data at $T$ = 150 K for 10 different $\theta$
values ranging from 0 to 45$^{\circ}$. For small $\theta$ (antinodal
region), the curves show two peaks at $\pm\omega_p$ symmetric to
$\omega$ = 0. As $\theta$ increases, the two peaks shift to smaller
$|\omega|$ and between $\theta$ = 35$^{\circ}$ and 40$^{\circ}$,
they merge into a single peak at $\omega$ = 0, thus forming the
Fermi arc as discussed in the ARPES experiment.
Fig.~\ref{fig:spectral}(b) shows the corresponding DOS from Eq.~(1),
in which the gapped structure disappears only at the nodal point of
$\theta$ = 45$^{\circ}$.

The ARPES gap $\omega_p$ can be found by setting the first
derivative of Eq.~(6) to zero:
\begin{equation}
\omega_p^2 =
(1+\frac{\Gamma_{\Delta}}{\Gamma})\Delta\cos(2\theta)\sqrt{\eta}-
\frac{\Gamma_{\Delta}}{\Gamma}\eta~,
\end{equation}
where $\eta$ = $\Delta^2\cos^2(2\theta)+\Gamma \Gamma_{\Delta}$. By
setting the second derivative to zero, the angle $\theta_0$ at which
the arc starts is found to be
\begin{equation}
\theta_0 =
0.5\cos^{-1}(\sqrt{\frac{\Gamma_{\Delta}}{\Gamma+2\Gamma_{\Delta}}}~
\frac{\Gamma_{\Delta}}{\Delta})~.
\end{equation}

\begin{figure}[t]
\includegraphics[width=0.42\textwidth]{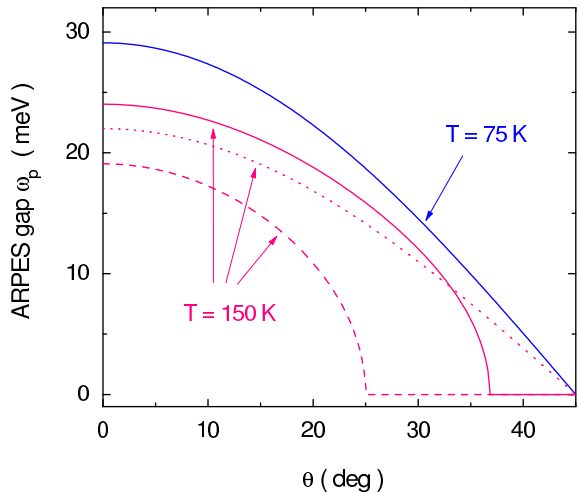}
\caption{(Color online) ARPES gap $\omega_p$ versus $\theta$ at $T$
= 75 and 150 K calculated from Eq.~(11) with $\Gamma$ and
$\Gamma_{\Delta}$ taken from the dashed lines with open symbols in
Fig.~\ref{fig:rates} (solid lines). The dashed line is the result at
$T$ = 150 K with equal $\Gamma$ = $\Gamma_{\Delta}$ = 24.5 meV. The
dotted line is the $d$-wave gap $\Delta\cos(2\theta)$ at $T$ = 150 K
shown for comparison.} \label{fig:arpesgap}
\end{figure}

Fig.~\ref{fig:arpesgap} displays $\omega_p$ as a function of
$\theta$ at two typical temperatures $T$ = 75 and 150 K (solid
lines). At $T$ = 75 K $<$ $T_c$, we have $\Gamma_{\Delta}$ = 0 and
from Eq.~(11), $\omega_p$ equals the full $d$-wave gap
$\Delta\cos(2\theta)$. At $T$ = 150 K, one can clearly see the
existence of the Fermi arc above $\theta_0$ $\sim$ 37$^{\circ}$. The
dotted line is the $d$-wave gap $\Delta\cos(2\theta)$ at the same
temperature plotted for comparison. In Fig.~\ref{fig:arpesgap}, we
also show the result at $T$ = 150 K under the special condition
$\Gamma$ = $\Gamma_{\Delta}$ = $\gamma_+$ (dashed line). The
corresponding $A(\theta,\omega)$ is displayed in the inset of
Fig.~\ref{fig:spectral}, which shows $\theta_0$ $\sim$ 25$^{\circ}$.
Setting $\Gamma$ = $\Gamma_{\Delta}$ at 150 K leads to the
$\Gamma_{\Delta}$ value approximately twice as large as that given
by Eq.~(10), as can be seen in Fig.~\ref{fig:rates}. In this case,
both ARPES gap and Fermi arc show considerable differences. In
Fig.~\ref{fig:arc}, the relative arc length $l_{arc}$, defined by
\begin{equation}
l_{arc} = 1-(\frac{4}{\pi})\theta_0~,
\end{equation}
is calculated from the $\Gamma$ and $\Gamma_{\Delta}$ parameters of
dashed lines with open symbols in Fig.~\ref{fig:rates}, and is
plotted as open symbols. The dashed line in the figure is a guide to
the eye, which extends to about 45 percent of the full Fermi surface
length near $T_1^*$.

\section{Discussion}

\begin{figure}[t]
\includegraphics[width=0.42\textwidth]{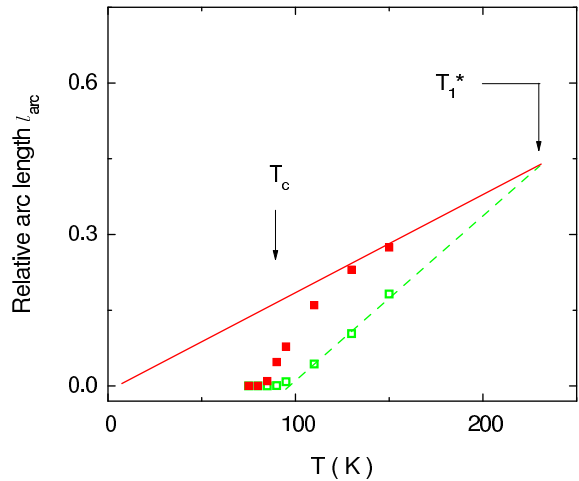}
\caption{(Color online) Relative arc length $\l_{arc}$ calculated
from Eq.~(13) using $\Delta$ in Fig.~\ref{fig:didvtc} and $\Gamma$
and $\Gamma_{\Delta}$ in Fig.~\ref{fig:rates}. Open and solid
symbols correspond to the $\Gamma$ and $\Gamma_{\Delta}$ parameters
shown as dashed line with open symbols and solid line with full
symbols in Fig.~\ref{fig:rates}, respectively. Two straight lines
are guides to the eye.} \label{fig:arc}
\end{figure}

Using Bi-2212 samples at various doping levels, Kanigel {\it et al.}
\cite{kani06} found the arc length above $T_c$ varying as
$T$/$T^{\star}$, extrapolating to zero as $T$ $\rightarrow$ 0, and
rapidly increasing from about 50 percent to the full Fermi surface
length near $T^{\star}$. This result is explained using the
self-energy model Eq.~(2) in the case of $\Gamma$ =
$\Gamma_{\Delta}$ and assuming a constant $\Delta$ and a linear
$\Gamma$ $\propto$ $T$ at all $T$. \cite{norm07} In the present case
with single doping sample, if we use Eq.~(10) to define
$\Gamma_{\Delta}$, we see that the arc length above $T_c$ has
approximately a linear temperature dependence and approaches about
45 percent of the full Fermi surface length near $T_1^*$, but it
extrapolates to zero as $T$ $\rightarrow$ $T_c$, as is shown by the
dashed line with open symbols in Fig.~\ref{fig:arc}. Such
temperature dependence of the arc length is similar to the
discussion in Ref.~\onlinecite{chub07} where the actual $\Gamma$
$\sim$ $T$ dependence from the ARPES measurements \cite{norm01} is
taken. However, it deviates from the experimental observations as
reported in Refs.~\onlinecite{kani06} and \onlinecite{kani07}.

In the present case of $\Gamma$ $\not=$ $\Gamma_{\Delta}$ we are
considering, we find that there exist alternative choices of the
$\Gamma$ and $\Gamma_{\Delta}$ parameters that can lead to a better
explanation. Instead of setting the $\Gamma_{\Delta}$ parameter
first from Eq.~(10), we can start in a reverse way by assuming a
$\Gamma$ linearly increasing with temperature, which extends from
the data below $T_c$ where $\Gamma_{\Delta}$ $\rightarrow$ 0. Such a
linear temperature-dependent quasiparticle scattering rate was
predicted in the marginal Fermi-liquid theory for the cuprate
superconductors \cite{varma89} and has been discussed in a number of
experiments. The $\Gamma$ parameter from this consideration is shown
in Fig.~\ref{fig:rates} as a solid line with full symbols, together
with the $\Gamma_{\Delta}$ parameter obtained from the
experimentally determined $\gamma_+$ through Eq.~(8). We can see
that the $\Gamma_{\Delta}$ parameter thus obtained exhibits a faster
rise with increasing temperature near $T_c$. The relative arc length
$l_{arc}$ calculated from Eq.~(13) using this set of parameters is
plotted in Fig.~\ref{fig:arc} as full symbols, which shows a clear
difference compared to that with open symbols. From the solid line
in the figure, one can see that the arc length would extrapolate to
zero as $T$ $\rightarrow$ 0, which is in an excellent agreement with
the ARPES experiment. \cite{kani06,kani07}

Experimentally, an ARPES gap opening on the Fermi arc as temperature
decreases right below $T_c$ was observed. \cite{shen07,kani07} The
gap opening can be gradual and follow the BCS-like temperature
dependence \cite{shen07} or has a sudden increase to the full
$d$-wave gap size. \cite{kani07} Our results, based on the tunneling
data and the self-energy model in Eq.~(2), seem to support the
latter observation provided $\Gamma_{\Delta}$ is zero or vanishingly
small below $T_c$ as is shown in Fig.~\ref{fig:rates}. Above $T_c$,
the Fermi arc exists and the $\theta$-dependence of the ARPES gap
significantly deviates from the $d$-wave gap behavior. One
interesting feature from our calculated results in
Fig.~\ref{fig:arpesgap} is that the ARPES gap for smaller $\theta$
at 150 K (solid line) surpasses that of the $d$-wave gap (dotted
line). Such a tendency may be compared to a recent ARPES measurement
on the La$_{1.875}$Ba$_{0.125}$CuO$_4$ material, \cite{he09} where a
clear evidence was found that near the antinodal region the ARPES
gap is enhanced and has a larger gap size compared to the $d$-wave
gap behavior near the nodal region. Possible origins of the
antinodal phenomenology including a gap opening with strong
quasiparticle scattering were mentioned in Ref.~\onlinecite{he09}.

The above discussion reveals a consistent picture from the key ARPES
observations and the corresponding results obtained from the
tunneling experiment through the phenomenological self-energy in
Eq.~(2). The consistency appears remarkable considering that only
$\theta$- and energy-independent parameters $\Gamma$ and
$\Gamma_{\Delta}$ are employed, and the only assumption used is a
linear temperature-dependent quasiparticle scattering rate $\Gamma$,
which should be reasonable for the cuprate superconductors.

It is worth emphasizing that the ARPES gap, from the description of
Eq.~(2), needs not always to be the energy gap $\Delta
\cos(2\theta)$. It can be zero (forming the Fermi arc) or larger
than $\Delta \cos(2\theta)$ depending on the values of $\Gamma$ and
$\Gamma_{\Delta}$. On the other hand, the gap parameter $\Delta$
from fitting the experimental spectra in Fig.~\ref{fig:didvtc}
(solid triangles) shows a distinct behavior compared to the BCS
temperature dependence (dashed and dotted lines). It decreases
quickly as $T$ approaches from below $T_c$ and there is a sudden
slope change at $T_c$. It sits slightly higher than that obtained
directly from the $dI/dV$ peak position (solid squares in the
figure) below $T_c$. Above $T_c$, their separation increases. In
Fig.~\ref{fig:didvtc}, one can also see that $\Delta$ tends to be
finite up to $T_1^{\star}$ =230 K, which in the precursor pairing
view suggests $T_1^{\star}$ to be the mean-field critical
temperature as discussed in the STM experiment. \cite{renn98}

The strange $\Delta$ behavior near $T_c$ could be caused by the loss
of the global coherence of Cooper pairs where the $\Delta$ parameter
should be understood as the thermodynamic-averaged value
$<\Delta^2>$. \cite{cohen69,tink96} In any case, the differences
between half the $dI/dV$ peak position, $\Delta$, and the system's
``true" superconducting gap parameter can be expected. At
temperatures above $T_c$, for example, the $dI/dV$ peak position
becomes largely affected by the increasing $\Gamma$ and
$\Gamma_{\Delta}$ (see Fig.~\ref{fig:didvtc}), as in the case of the
BCS superconductors. \cite{cohen69} Also, our $d$-wave Eliashberg
analysis \cite{zhao07} considering a bosonic spectrum
\cite{zasa,zhao07} has resulted in a satisfactory fit for the 4.2 K
curves in Fig.~\ref{fig:expiv} and a low-$T$ gap parameter
$\Delta_0$ $\sim$ 38 meV as defined from the gap function
$\Delta(\omega)$ by $\Delta_0$ = $\Delta(\Delta_0)$. \cite{schr64}
This $\Delta_0$ value, corresponding to the peak position of the
antinodal DOS from the theory, is greater than that obtained from
the $dI/dV$ peak position ($\sim$ 32 meV, see
Fig.~\ref{fig:didvtc}), which gives rise to
2$\Delta_0$/$kT_1^{\star}$ $\sim$ 3.83 close to the BCS $d$-wave
ratio of 4.28.

\section{Conclusion}

Tunneling spectra of near optimally doped, submicron
Bi$_2$Sr$_2$CaCu$_2$O$_{8+\delta}$ intrinsic Josephson junctions
were presented, and examined in the region near and above $T_c$
where the superconducting gap evolves into pseudogap. We showed that
the spectra can be well fitted using the density of states in the
familiar Dynes' form, which is closely related to a self-energy
model proposed recently to discuss the Fermi arc phenomena observed
in the ARPES experiment. In this work, we considered a general
situation where the quasiparticle scattering rate $\Gamma$ and pair
decay rate $\Gamma_{\Delta}$ in the model are independent
parameters, related however through $\gamma_+$ = $(\Gamma +
\Gamma_{\Delta})$/2. We compared the Fermi arc behavior in the
pseudogap phase from the tunneling and ARPES experiments, and showed
that from the experimentally determined $\gamma_+$ parameter, some
key ARPES results can be derived based upon only one assumption: a
linear temperature-dependent $\Gamma$. This demonstrates a
remarkable consistency between the two experiments, which is in
favor of the precursor pairing view of the pseudogap. \\

We are grateful to L. Y. Zhang, T. Xiang, N. L. Wang, and Q. H. Wang
for many valuable discussions. We thank M. R. Norman for pointing
out Ref. \onlinecite{norm98a} to us. This work was supported by the
Ministry of Science and Technology of China (2006CB601007), the
Knowledge Innovation Project of the Chinese Academy of Sciences, and
the National Natural Science Foundation of China (10604064).



\begin{thebibliography}{}
\bibitem{rem1}Clear dip features, resulting from current overshoot
above the $R_N$ line near 80 mV, were observed in the submicron IJJs
from over 20 crystals fabricated.
\end{thebibliography}
\end{document}